\documentclass[proof]{pasj00}
\Received{$\langle$reception date$\rangle$}
\Accepted{$\langle$acception date$\rangle$}
\Published{$\langle$publication date$\rangle$}
\SetRunningHead{Astronomical Society of Japan}{Pre-flare characteristics}

\usepackage{times}

\begin{document}

\title{Coronal Behavior Before the Large Flare Onset}

\author{ Shinsuke Imada,\altaffilmark{1}
Yumi Bamba,\altaffilmark{1}
Kanya Kusano\altaffilmark{1,2}}
\altaffiltext{1}{ Solar-Terrestrial Environment Laboratory (STEL), Nagoya University, Furo-cho, Chikusa-ku, Nagoya 464-8601, Japan}
\altaffiltext{2}{Japan Agency for Marine-Earth Science and Technology, Yokohama, Kanagawa, Japan }
\email{shinimada@stelab.nagoya-u.ac.jp}


\maketitle

\begin{abstract}
Flares are a major explosive event in our solar system.
They are often followed by coronal mass ejection that has a potential to trigger the geomagnetic storms.
There are various studies aiming to predict when and where the flares are likely to occur. 
Most of these studies mainly discuss the photospheric and chromospheric activity before the flare onset. 
In this paper we study the coronal features before the famous large flare occurrence  on December 13th, 2006. 
Using the data from {\it Hinode}/EUV Imaging Spectrometer (EIS), X-Ray Telescope (XRT), and {\it Solar and Heliospheric Observatory} ({\it SOHO}) /Extreme ultraviolet Imaging Telescope (EIT), we discuss the coronal features in the large scale (~ a few 100 arcsec) before the flare onset. 
Our findings are as follows: 
1) The upflows in and around active region start growing from $\sim$10 to 30 km s$^{-1}$ a day before the flare. 
2) The expanding coronal loops are clearly observed a few hours before the flare.
3) Soft X-ray and EUV intensity are gradually reduced. 
4) The upflows are further enhanced after the flare.
From these observed signatures, we conclude that the outer part of active region loops with 
low density were expanding a day before the flare onset, and the inner part with high density were expanding a few hours before the onset. 
\end{abstract}

\section{Introduction}
A sudden brightening of the solar surface in optical wavelength was observed by Richard Carrington for the first time in 1859 (\cite{carr}). This sudden brightening, which is now called a solar flare is believed to be the result of rapid release of magnetic energy stored in the solar corona. 
The energy released by a flare is very large and the total amount of this energy often reaches $10^{32}$ ergs within an hour. The brightening released by the flare is observed in almost all wavelengths from radio waves to gamma rays.
Solar flares are often associated with coronal mass ejections (CMEs) which can trigger geomagnetic storms.
Therefore, it is crucial to understand the physical mechanisms of eruptive flares from the view point of solar physics as well as space weather. 

The mechanism of eruptive solar flares is a long standing problem after its discovery.
Magnetic reconnection was proposed as a physical mechanism behind the eruption of flares nearly 100 years after the discovery of Carringthon flare event (\cite{par, swe,pet}). 
Magnetic reconnection can rapidly release the magnetic energy stored inside the current sheet in the form of  thermal energy, kinetic energy or non-thermal particle energy. 
These energy conversions are fundamental and essential in understanding the dynamical behavior of solar corona. 

It is important to understand what magnetic configuration can drive the magnetic reconnection. 
Considerable effort has been devoted toward constructing the physical model of eruptive flares, and several models have been proposed. 
One standard model of eruptive flares that is based on magnetic reconnection is the CSHKP model (\cite{car,stu,hir,kop}). There are extended models based on the CSHKP model (e.g., \cite{for,aul}). The predicted characteristics from these models have been verified by modern observations (e.g., cusp-like structure in soft X-ray images: \cite{tsu}, hard X-ray source above the flare loop: \cite{mas}, chromospheric evaporation: \cite{ter}, reconnection inflows: \cite{yok}, reconnection outflows (off limb: \cite{mck,inn,ima6}, on disc: \cite{har}), plasmoid ejection: \cite{ohy,liu}, and CMEs: \cite{sve}). 
However, the detailed mechanisms that drives the system into an eruptive stage, and eventually results in the eruption are not yet fully understood.

Many studies consider the observations of magnetic activities at photosphere just before the onset of eruptive flares.
These magnetic structures include strong magnetic shear (e.g., \cite{low,hag}), reversed magnetic shear (e.g., \cite{kus,wan2}), flux cancelation (e.g., \cite{van,zha}),  converging foot point motions (e.g., \cite{inh,for2}), sharp gradient of magnetic field (e.g., \cite{sch}), emerging magnetic flux (e.g., \cite{hey,che}).
In fact, there have been emphasis on causal relationship between these magnetic properties and the occurrence of solar eruptions.
However, most of these studies discuss the flare trigger through the photospheric/chromospheric magnetic field structures before the flare onset.
Few studies look at the plasma conditions at the solar corona before the flare eruption.
\cite{harr} discussed the pre-flare behavior of the corona a few days before the large solar flare on 2006 December 13.
They examined the response of the corona to the helicity injection and found that an increase in the coronal spectral line widths is observed after the time saturation of the injected helicity is measured (\cite{mag}).
Line broadening of coronal spectra is believed to be a consequence of nonthermal motions. 
This increase in line widths happens before any eruptive activity.
Therefore, there might be a possible causal relationship between line broadening in EUV wavelengths and the flare onset.

In this paper, we discuss the coronal features before the famous large flare occurrence on December 13th, 2006. 
We use data from {\it Hinode}/EIS, XRT, and {\it SOHO}/EIT  to discuss the coronal features in the large scale before the flare onset. 
Specifically we are interested in the relationship between the nonthermal line broadening in EUV wavelength and the flare onset, and we examine the results of  \cite{harr}. We discuss the origin of the nonthermal line broadening before the flare  onset.

\section{Observations and Data Analysis}
\subsection{Instrumentation}
The Hinode spacecraft was launched on the 22nd September 2006 (\cite{kos}). 
It is a Japanese mission with collaborations from the US and UK and with three instruments onboard (the solar optical telescope (SOT), XRT and EIS). The EIS instrument  onboard {\it Hinode} is a high spectral resolution spectrometer aimed at studying dynamical phenomena in the corona with high spatial resolution and sensitivity (\cite{cul}). 
In this section,we discuss the three sequences of EIS observations from 01:07 UT on 12th December 2006 to 05:42 UT on 13th December 2006. All EIS data used in this paper are the same raster scanning observations (JTM004).
JTM004 contains nine spectral windows that can cover the wide temperature range $4.9<$ logT $<6.7$. 
The slit width and the scanning steps are both 1 arcsec, and the exposure time is 30 sec.
It takes $\sim$ 4.5 hour to obtain one scanning image (512$\times$256 arcsec$^2$).
EIS data from the raster are processed using the EIS team software ({\it eis\_prep}), which corrects for the flat field, dark current, cosmic rays, and hot pixels. The slit tilt was corrected by {\it eis\_tilt} correction.
For thermal reasons, there is an orbital variation of the line position causing an artificial Doppler shift of +/- 20 km s$^{-1}$ which follows a sinusoidal behavior. This orbital variation of the line position was corrected using the house keeping data (\cite{kam}).
While making a raster scan from 01:12 UT to 05:42 UT on 2006 December 13, EIS observed an X3.2 flare.
The flare occurred at 02:14 UT accompanied by a halo CME.
The flare itself has already been studied in detail (e.g., \cite{ima2, asa,min}).

For studying the temporal evolution of coronal activity before the flare onset, we made use of the EIT data (\cite{del}). 
We used the EIT images in the 195  \AA~ filter. 
The EIT data has a cadence of 12 mins and a pixel size of 2.6 arcsec pixels. 
The exposure times are $\sim$12.6 sec during the event. 
We processed EIT data by using eit\_prep for the calibration.

\subsection{Results}

\subsubsection{Imaging Observations by SOHO/EIT}
In order to study the temporal evolution of corona before the flare onset, we analyze the EIT data from 2006 December 12th 23:49 to 13th 5:48 UT. We study the temporal evolution of the active region of NOAA 10390. 
The temporal variation of the active region in 195 \AA~ filter can be seen in the movie (Mov\_1.mov). 
This movie provides a clear impression of the expanding coronal loops just before the flare onset.
Figure 1 shows the temporal evolution of coronal loops that are located on the east side of the active region before and after the X-class flare. The white dotted lines show the direction for loop expansion, and the white horizontal arrow represents the top of the expanding loop. The white vertical arrow shows the footpoint position of the expanding loop.
As it can be seen in Figure 1, the coronal loop gradually expands before the flare.
The flare occurred between 02:12 and 02:25 UT, although the EIT image is saturated after the flare onset.
The expanding loop vanished just after the flare (02:36 UT).
The inclination of the loop at the foot point which is marked by vertical white arrow in Figure 1 becomes steeper with time.
This result could indicate that the loop height increases with time.
Figure 2 shows the temporal evolution of coronal loops by using the running difference technique.
Each EIT image is subtracted by the EIT image obtained at the previous time.
The white arrows and dotted lines are the same as in Figure 1.
The expanding loop can be seen more clearly in Figure 2, because the outline of coronal loop is emphasized by using running difference technique.
It is also seen in the last panel of Figure 2 that the expanding loop vanished just after the flare onset (black represent vanishing).  
To estimate the apparent velocity of the expanding coronal loop, we create a time-distance plot  along the white dotted line in Figure 1. The time-distance plot in Figure 3 shows the velocity with a typical value of $\sim10$ km sec$^{-1}$, which is sufficiently slower than the sound speed and Alfv\'en speed in the solar corona.   

Figure 4 shows another example of the expanding loop before the flare onset.
The coronal loop of interest is located in the southern part of the flare productive active region.
The loop gradually expands along the white dotted line before the flare onset.
The inclination of the footpoint shows the same behavior as in Figure 1.
Figure 5 shows the temporal evolution of coronal loops by using running difference technique, and the expanding loop is seen clearly in Figure 5.
The last two panels of Figure 5 show that the expanding loop vanished just after the flare onset.  
The estimated velocity of this loop which is based on the time-distance plot (Figure 3) is $\sim 10$ km sec$^{-1}$. 

The coronal loops overlaid above the core of active region indicated by red arrows vanished a few hours before the flare onset.
Figure 6 shows the EIT image zooming up to the active region on the 12th December 23:49 and 13th December 01:12 UT. 
The white arrows indicate the location of the overlaid loops above the active region.
We can see that the overlaid loops vanished between 12th at 23:49 and 13th at 01:12 UT.
Figure 7 shows the soft X-ray image obtained by {\it Hinode}/XRT two hour and ten minutes before the occurrence of flare.
At 02:04 on 13th December UT (right panel of Figure 7) we can see the sigmoidal structure in the core part of the active region.
However, two hours before, at 23:50 on 12th December UT (left panel of Figure 7) the sigmoidal structure cannot be seen because of the overlaid coronal loops. Figure 8 shows the temporal evolution of the total soft X-ray/EUV obtained by {\it GOES} (top), {\it Hinode}/XRT (middle), and {\it SOHO}/EIT (bottom), respectively.
The soft X-ray light curve obtained by GOES represents the X-ray not only from the flare productive active region but also from the whole sun. The active region of NOAA 10930 discussed in this paper is the only active region at that time.
Thus, GOES light curve represents the characteristics of this flare productive region.  
The soft X-ray and EUV light curves obtained by XRT and EIT are averaged by the area which is shown in Figures 6 and 7.
The middle and bottom panel in Figure 8 shows the light curve of the flare productive active region.
Note that there is a data gap in the XRT data from $\sim$00:45 to $\sim$02:00 UT.
From Figure 8 we can conclude that the coronal emission in soft X-ray and EUV wavelengths reduces before the flare onset.

\subsubsection{Spectroscopic Observations by Hinode/EIS}
EIS observed the active region NOAA 10930 with three sequences of raster scans from 01:07 UT on 12th December 2006 to 05:42 UT on 13th December 2006.
Figure 9 shows the intensity, velocity, and line width map of Fe\emissiontype{XII} (195.12 \AA; $\log T_{max} = 6.11$) before and after the flare onset. While making the last raster scan from 01:12 UT to 05:42 UT on 2006 December 13,  an X3.2 flare occurred (02:14 UT).
The color scales in the velocity map range from -30 to 30 km sec$^{-1}$, and the color scales in the line width map range from 40 to 120 km sec$^{-1}$. Both color scales are fixed in the three sequences of observations.
The line width includes the instrumental width, thermal width, and the nonthermal width.
Note that EIS scans from west (right) to east (left) with $\sim$4.5 hour.
The times displayed in Figure 9 represent the start time of each scanning. 

The active region is located at $(X,Y)\sim (150,-100), (300,-100), (350,-100)$, from the top to bottom in Figure 9, respectively.
The plage region is located 200 arcsec east from the active region.
A day before the flare, as is shown in Figure 9b, the doppler velocities  are $<5$ km sec$^{-1}$ in most part of the active region.
At the northern boundary of active region, we can see relatively strong blueshifts ($\sim 10$ km sec$^{-1}$) in Figure 9b.  
These upflows might be related to the outflows observed at the edge of usual active regions (e.g., \cite{sak,harr2}). 
We do not observe any typical pre-flare feature in doppler velocity (Figure 9b) or line width (Figure 9c) a day before the flare.
A few hours before the flare, as is shown in Figure 9e, the doppler velocity is enhanced by up to $\sim 30$ km sec$^{-1}$ around the active region. This result means that the upflows around this region are growing from $\sim 10$ to 30 km sec$^{-1}$ a day before the flare. Further we find that these upflows are correlated with the line broadening (Figures 9(e-f)).
The line widths in the blueshifted regions almost reach 100 km sec$^{-1}$.
We have put three squares around the typical blueshifted region in Figures d-f.
We will discuss the line profiles inside the square region later. 
The plage region which is located 200 arcsec east from the active region slightly change its characteristics.
Because we can observe the upflows and the reduction of intensity, the plage region might become transient coronal hole. 
Actually the coronal dimming is observed at the plage region after the flare, and it is believed to be one of the source of the CMEs (e.g., \cite{ima, ima4}). 
As Figure 9h shows, after the flare onset or just before the onset the blueshifted regions are further growing.
The doppler velocity in some parts reach $>50$ km sec$^{-1}$, and the line width reaches 120 km sec$^{-1}$.
The distribution of the strong blueshifted regions in Figure 9h (after the flare) seem to be associated with the blueshifted regions of Figure 9e (before the flare).
Therefore, the blueshift observed a few hours before the flare onset might be a typical signature of pre-flare activity.

Figure 10 shows examples of the line profile observed at the blue shifted region in and around the active region (Figures 10(a-c)).
Enlarged displays of the EIS observations which are framed by squares in Figure 9d-f are shown (Figures 10(d-f): intensity, (g-i): velocity, (j-l): line width).
White contours represent the blueshift faster than 20 km sec$^{-1}$.
Figures 10(a-c) show examples of the Fe\emissiontype{XII} line profile of the blueshifted regions (marked by x in Figures 10(d-l)) with the Gaussian fit overlaid. The line profiles in Figure 10(a)/(b)/(c) correspond to the x-marks in Figures 10(d, g, j)/(e, h, k)/(f, i, l), respectively. Every line profile is almost symmetric and its center is blueshifted, indicating the plasma flow upward.
The blueshifted regions can be observed at the edges of the active region or plage regions where the intensities are relatively low compared with the rest of the active region. We can see clearly that the line widths are also large in the blueshifted regions.

\section{Summary and Discussion}
We have provided a comprehensive study of the coronal features before the large flare using observations by {\it Hinode} and {\it SOHO}. Our findings are as follows; 
\begin{itemize}
\item A day before the flare the upflows in and around the active region are growing from $\sim 10$ to $30$ km sec$^{-1}$.
\item These upflows  are clearly correlated with the line broadening.
\item A few hours before the flare
the coronal loops in and around the active region expand by $\sim 10$ km sec$^{-1}$.
\item A few hours before the flare the intensity of the active region in soft X-ray or EUV wave length is gradually reduced. 
\item After the flare the observed upflows in and around the active region are further growing ($>50$ km sec$^{-1}$).
\end{itemize}
These observed signatures indicate that the outer part of the active region start to expand a day before the flare (1st stage). 
Because the density of the outer part of the active region is low,  only the doppler velocity observations can detect their expanding signature. 
In the inner part of the active region the density is relatively high, and it starts to expand a few hours before the flare (2nd stage). 
At this stage the expanding structure can be observed with EIT.
The overlaid coronal loops above the core of the active region also start to expand a few hours before the flare.
Figure 11 shows the schematic illustration of the observed coronal features before the large eruptive flare.
The coverage in time of our observation is limited because of the low temporal resolution in EIS observation.
We need high cadence observation to confirm our scenario precisely.
This is important for future work.

A linear expansion of coronal loops or prominence have been reported several times (e.g., \cite{ste2}).
\cite{ste3} and \cite{chi} discussed the early phase of prominence eruption by using the EUV imaging observations.
They found that there are two typical phase for filament eruption.
One is the slow-rise phase ($\sim$10 km sec$^{-1}$), and the other is fast-rise phase (a few 100 km sec$^{-1}$).   
The relationship between these two phase and the initiation of coronal mass ejection is intensively discussed (e.g., \cite{moo}).
It is now believed that the coronal loops reconfiguration in large scale occurred and some loops become taller and/or longer before the main phase of the eruption.
Therefore, our interpretation (Figure 11) is consistent with the past observations.

The outflows observed at the edge of active regions have been reported several times not only imaging but also spectroscopic observation in EUV (e.g., \cite{sak,harr}).
The typical outflow velocities are $\sim$  a few 10 km sec$^{-1}$, and their temporal variations are seems to be small (e.g., \cite{dem}). 
There are two types of the models to explain the origin of active region outflows.
One is the wave-induced flow models (e.g., \cite{wan}), and the other is  reconnection-induced flow models(e.g., \cite{bak}).
We found that the outflows in and around the active region are growing before the flare, although active region flows are generally steady.
Therefore, we think that the loop expansion effect, in addition to wave and/or reconnection effect, is the key to understand the growth of active region outflows before the flare.

As we mentioned in the Introduction, \cite{harr} studied the causal relationship between the nonthermal broadening in EUV wavelengths and the flare onset by using the same flare event. 
Further, they found that the two pre-eruptions are observed before the flare (December 12, 16:40 and 23:00) and claimed that the increase in line widths start before any eruptive activity occurs. 
Our observations shows a clear relationship between the upflows and the line broadening. 
It is often observed that the line broadening is related to the flow (e.g., \cite{dos}).
Therefore, the origin of the line broadening before the flare might be related to the upflows.
The changing magnetic configuration by loop expansion causes the evacuation of coronal material toward higher corona through the rarefaction wave. 
Thus, the loop expansion causes dimming and upflows which velocity is slower than the sound and Alfv\'en speed in the solar corona.
We think that the origin of the nonthermal line broadening that is observed before the flare is the flow during the loop expansion.

\cite{ste} also discussed the relationship between the two pre-eruptions and the main eruption that are discussed in \cite{harr}. They found that there are lateral offset between them. In our observations, we observe the loop expansion in and around the active region before the flare. There are also lateral offset between the core active region loops and expanding loops.
Our result is quite similar to their result in the sense that there are lateral offset between the slow expanding loops and main flaring loops. 

The relationship between flares/reconnection and CME has been discussed for a long time.
A lot of models have been proposed to explain the relationship (e.g., \cite{moo}).
It is still controversial whether flares cause CMEs or vice versa.
Recently, \cite{kus2} studied the flare triggering process by using the three-dimensional magnetohydrodynamic (MHD) simulations.
They systematically examined the simulations in a wide variety of magnetic configurations.
They discussed the relationship between the two type of flare,  "reconnection-induced eruption" and "eruption-induce reconnection", and the characteristics of triggering magnetic field.
They claimed that the triggering magnetic fields can be devided into two types.
The opposite-polarity (OP) and reversed-shear (RS) type (see \cite{kus2} for detail). 
They state that in the case of OP-type flare "eruption-induced reconnection" flare might occur. 
In our observations the reconfiguration of magnetic fields seem to start before the flare, because they are continuously expanding from a day before the flare. 
Therefore, it is plausible that the flare discussed in this paper is "eruption-induced reconnection" flare.
\cite{bam} studied the flare triggering process by using the {\it Hinode} photospheric/chromospheric observations.
They claim that the flare on 2006 December 13 was OP-type.
Further, the flare triggering process seems to start after $\sim$00:00UT.
The coronal loop expansion observed by EIT in our observation also starts after $\sim$00:00UT.
Thus, our results based on the coronal features before and after the eruption is consistent with "eruption-induced reconnection" flare scenario discussed in \cite{kus2} and \cite{bam}. 
It is important to test our results in other flares that "reconnection-induced eruption" flare. This is what we are intending to do as part of our future work. 

Recently, \cite{ima5} showed that the density reduction in global scale is important to the onset of fast magnetic reconnection.
The transition of magnetic reconnection from collisional to collisionless can be achieved by density reduction.
An important future step would be to discuss the changing of plasma condition through the coronal loop expansion and its impact on fast magnetic reconnection.

Finally, we state the flare predictions to which our results may apply.
We believe that the key element of pre-flare is loop expansion in and around the flare productive active region.
From our results the loop expansion signature might be observed by the EUV spectroscopic observation a day before the eruptive flare.
The EUV Imaging observations may also be able to observe the signature a few hours before the flare.
If our scenario is correct, we can discuss the flare prediction with two kind of period (a day and a few hours).
We believe that our result is helpful to improve the prediction and probability of the flare occurrence.

\section{Acknowledgment}
The authors thank M. Asgari-Targhi and T. Magara for fruitful discussions.
Hinode is a Japanese mission developed and launched by ISAS/JAXA, collaborating with NAOJ as a domestic partner, NASA and STFC (UK) as international partners. Scientific operation of the Hinode mission is conducted by the Hinode science team organized at ISAS/JAXA. This team mainly consists of scientists from institutes in the partner countries. Support for the post-launch operation is provided by JAXA and NAOJ (Japan), STFC (U.K.), NASA (U.S.A.), ESA, and NSC (Norway).
This work was partially supported by the Grant-in-Aid for Young Scientist B (24740130), by the Grant-in-Aid for Scientific Research B (23340045), by the Grant-in-Aid for Scientific Research B (26287143), by the JSPS Core-to-Core Program (22001).

\begin{figure}
\begin{center}
\FigureFile(150mm,80mm){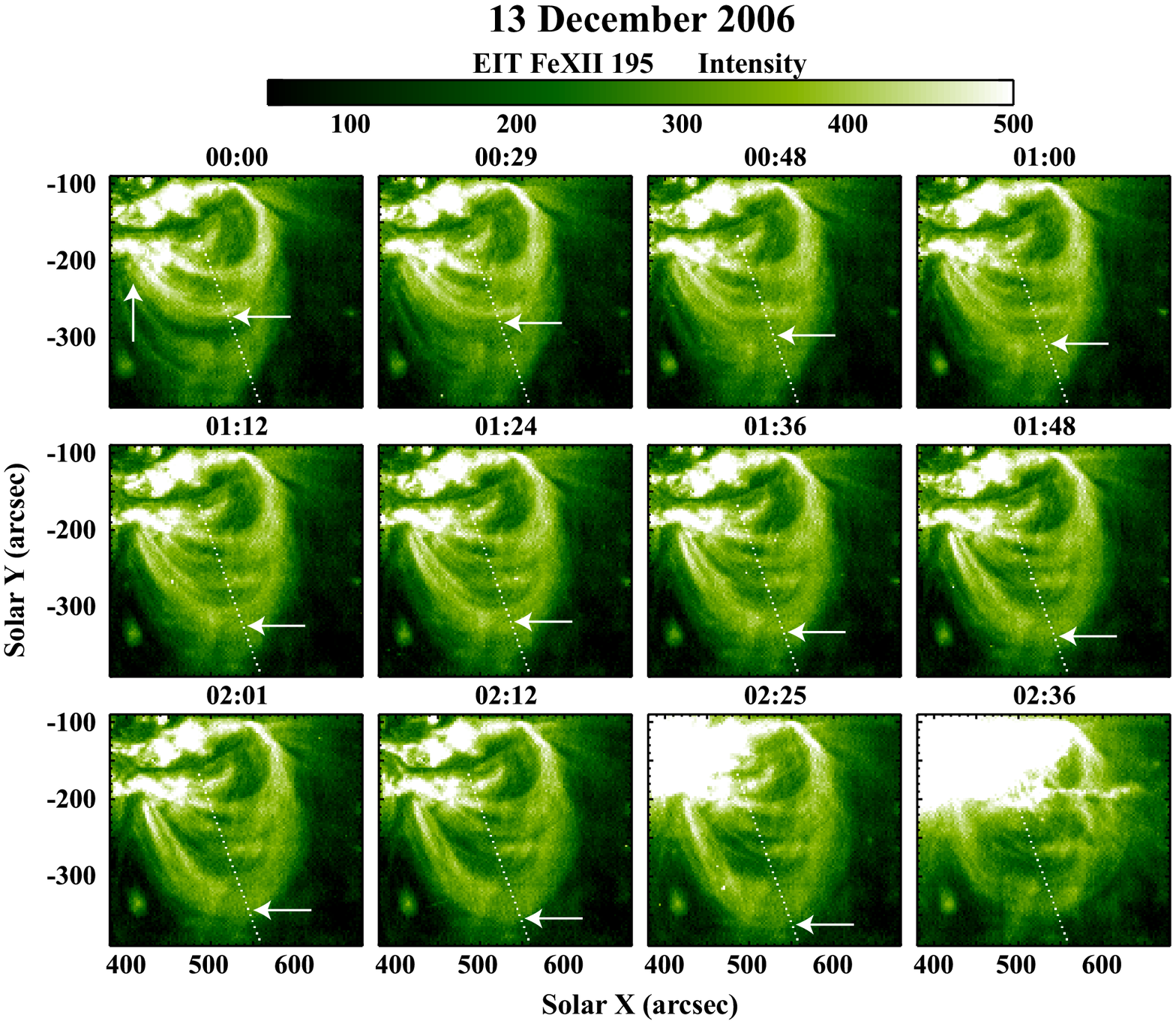}
\end{center}
\caption{ Temporal evolution of coronal loops located in the  east side of the active region observed by EIT (195 \AA~) before and after the large flare. The white arrow shows the top of the loop, and the white dotted line shows the direction of the loop expansion. 
The flare occurred at 02:14 UT. 
}\label{test}
\end{figure}

\begin{figure}
\begin{center}
\FigureFile(150mm,80mm){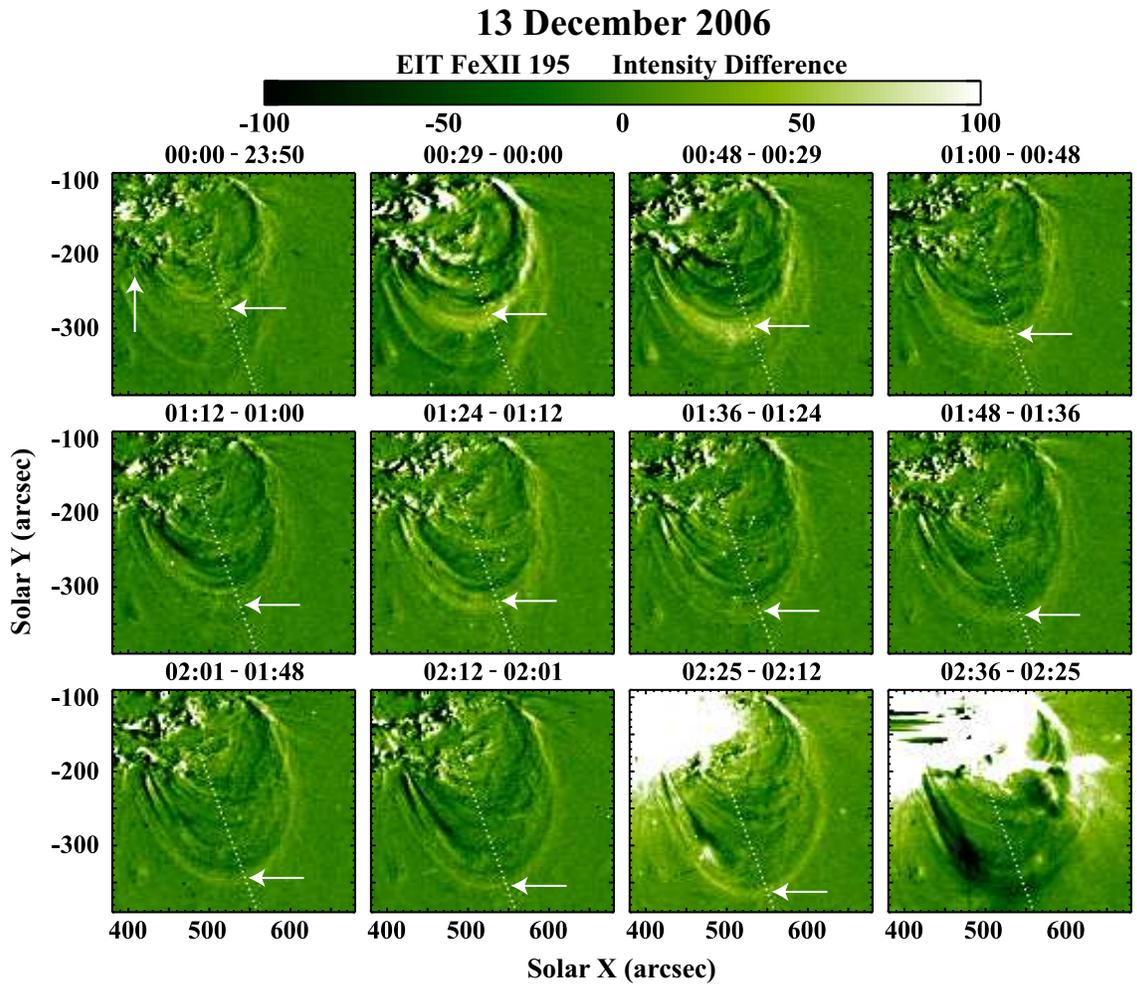}
\end{center}
\caption{ The emphasized figure of the temporal evolution of the coronal loops shown in Figure 1. This image is created using the running difference technique.
}\label{test}
\end{figure}

\begin{figure}
\begin{center}
\FigureFile(150mm,80mm){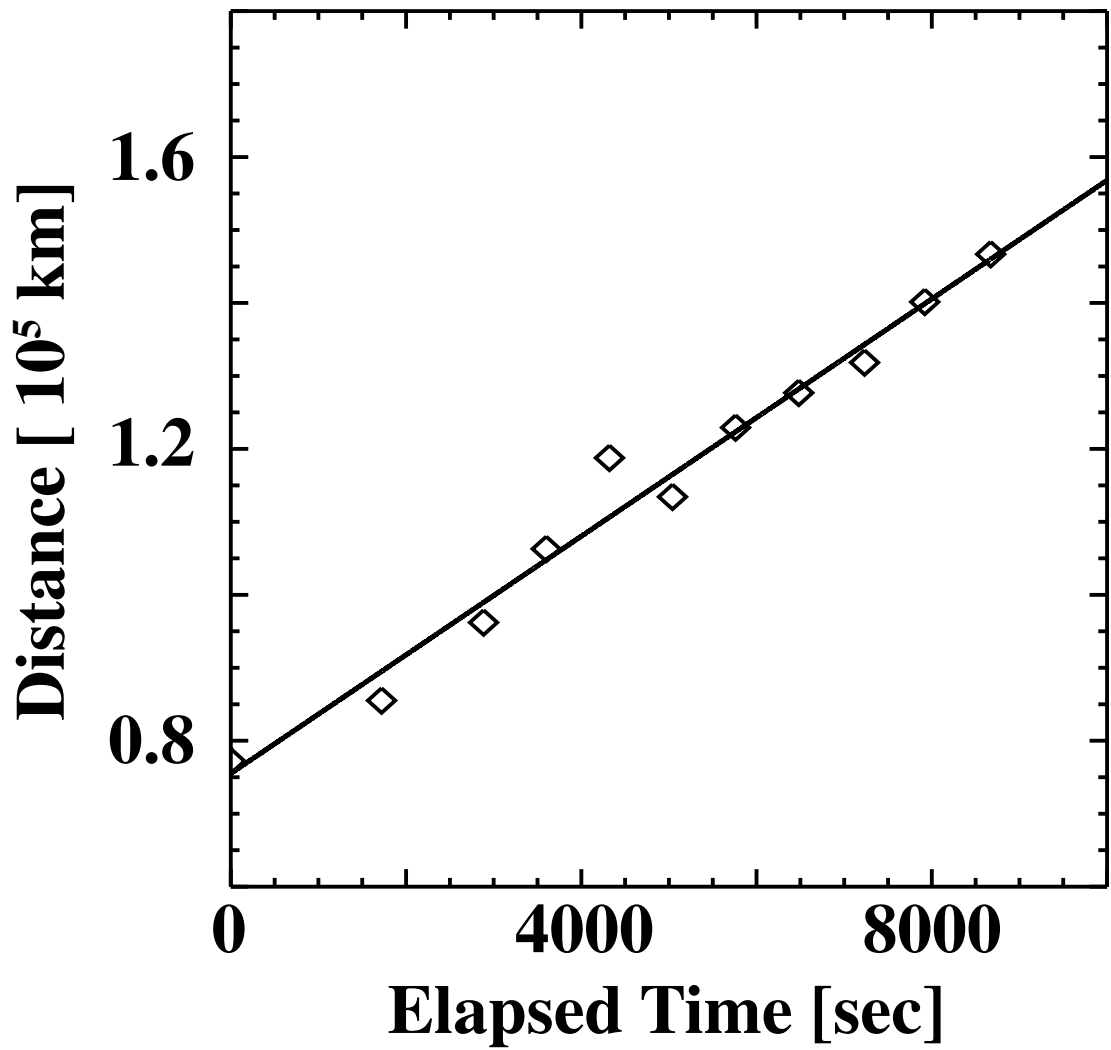}
\end{center}
\caption{ The time-distance plot along the white dotted line of Figure 1. The diamonds represent the height of the loop top, and the solid line shows the linear fitting result ($\sim 10$ km sec$^{-1}$).
}\label{test}
\end{figure}

\begin{figure}
\begin{center}
\FigureFile(150mm,80mm){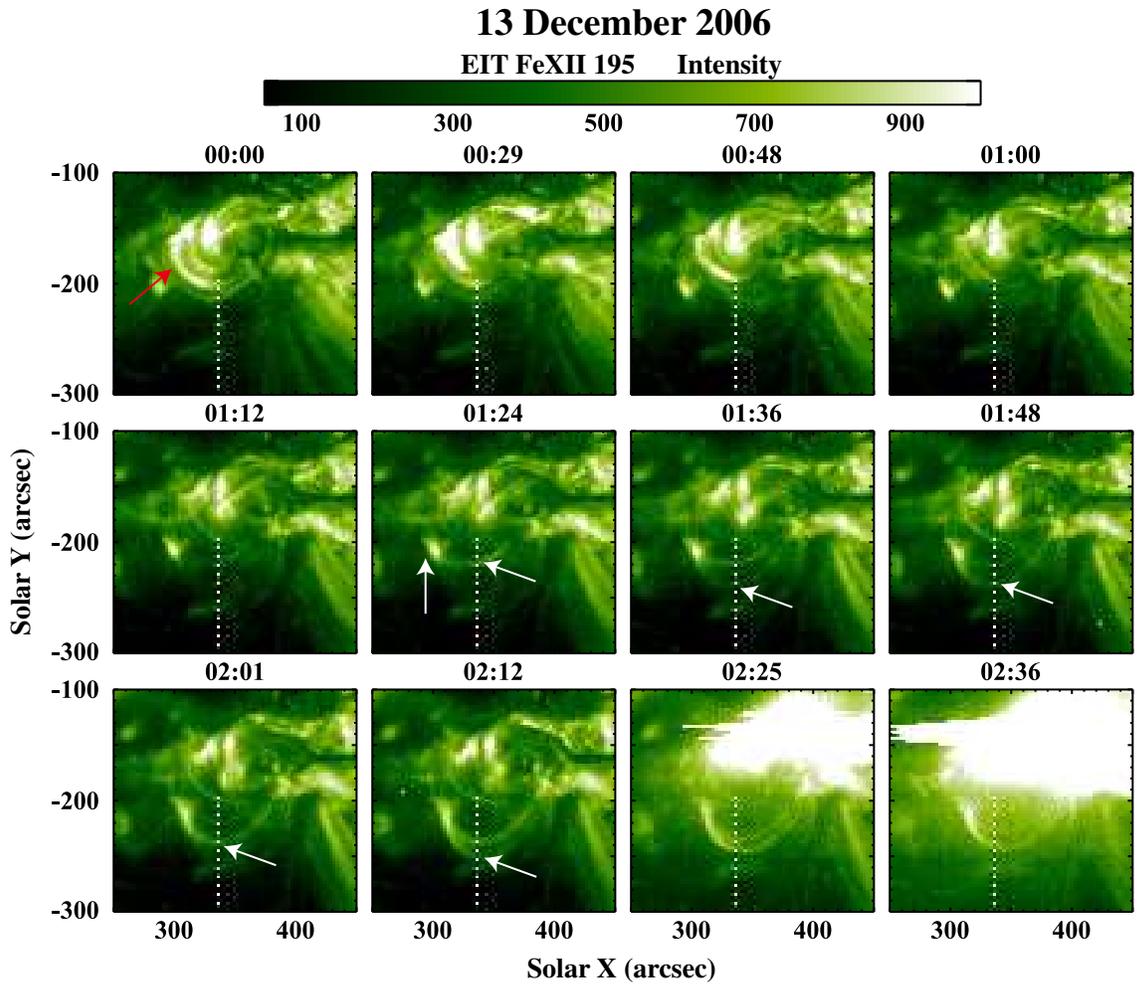}
\end{center}
\caption{ Another example of the expanding loop before the flare onset. The figure structure is the same as Figure 1. 
Red arrow shows the coronal loops overlaid above the core of the active region.
}\label{test}
\end{figure}

\begin{figure}
\begin{center}
\FigureFile(150mm,80mm){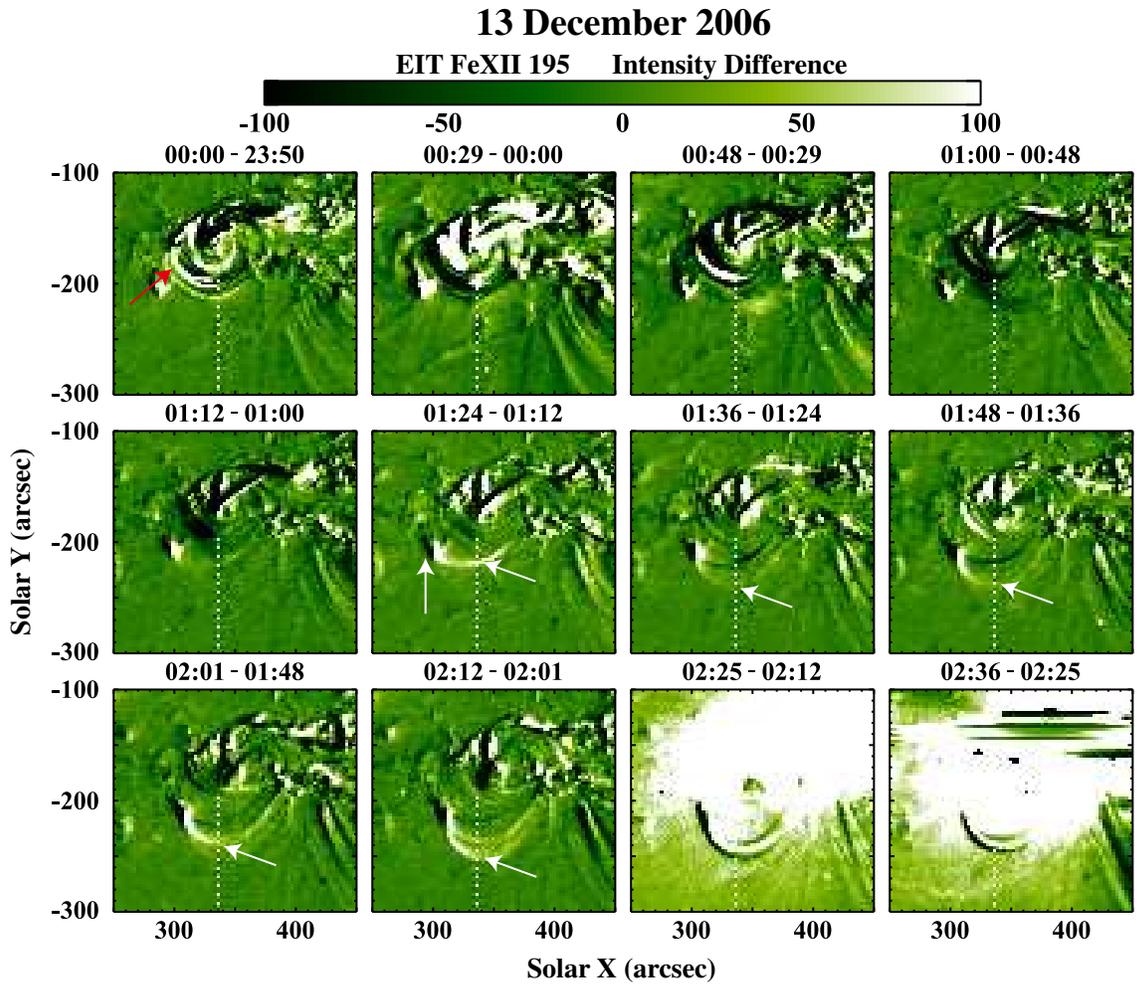}
\end{center}
\caption{ The emphasized figure of the temporal evolution of coronal loops shown in Figure 4 by using the running difference technique.
}\label{test}
\end{figure}

\begin{figure}
\begin{center}
\FigureFile(150mm,80mm){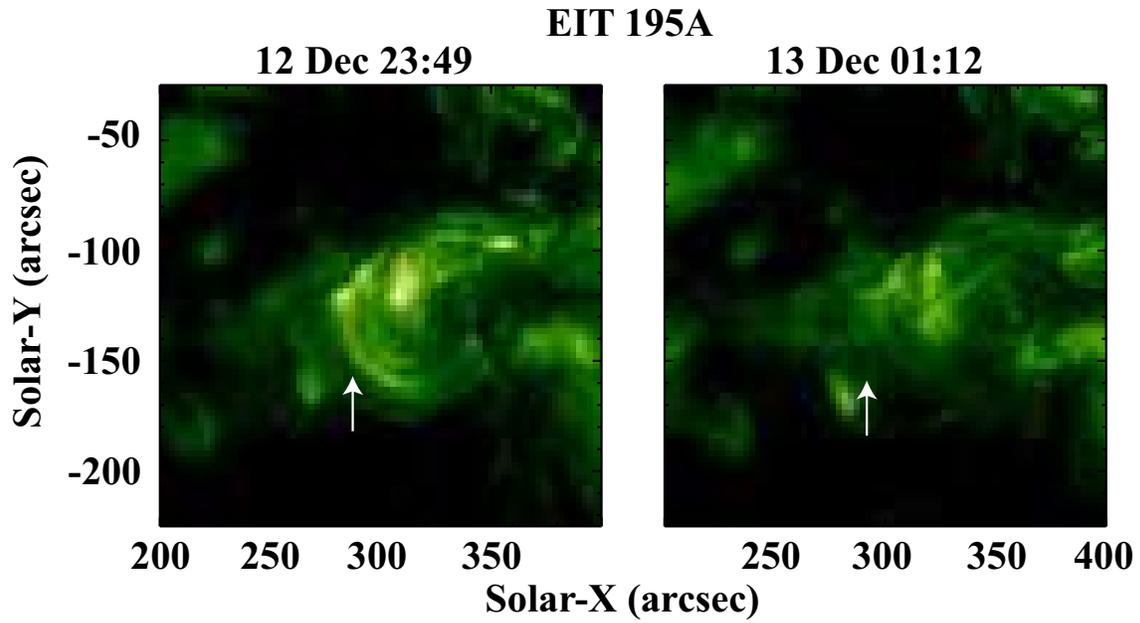}
\end{center}
\caption{ Enlarged display of the active region observed by EIT in 195 \AA~  at 23:49 on 12th (left) and at 01:12 on 13th  (right) December UT. The white arrows show the coronal loops overlaid above the core of active region.
}\label{test}
\end{figure}

\begin{figure}
\begin{center}
\FigureFile(150mm,80mm){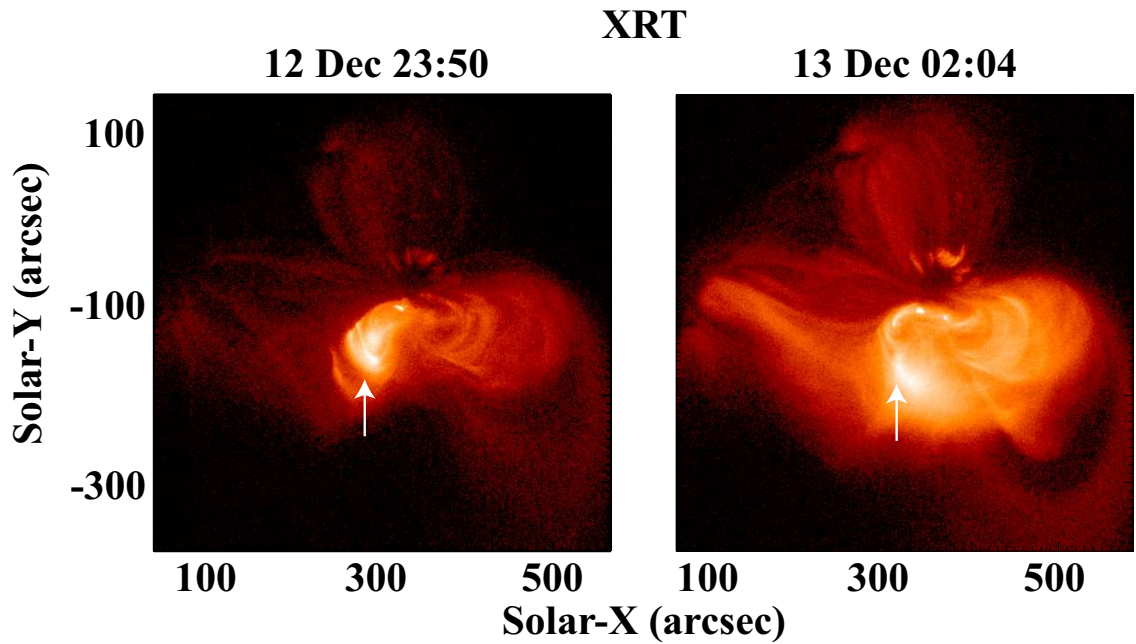}
\end{center}
\caption{ Soft X-ray image obtained by {\it Hinode}/XRT  at 23:50 on 12th (left) and at 02:04 on 13th  (right) December UT.
}\label{test}
\end{figure}

\begin{figure}
\begin{center}
\FigureFile(80mm,80mm){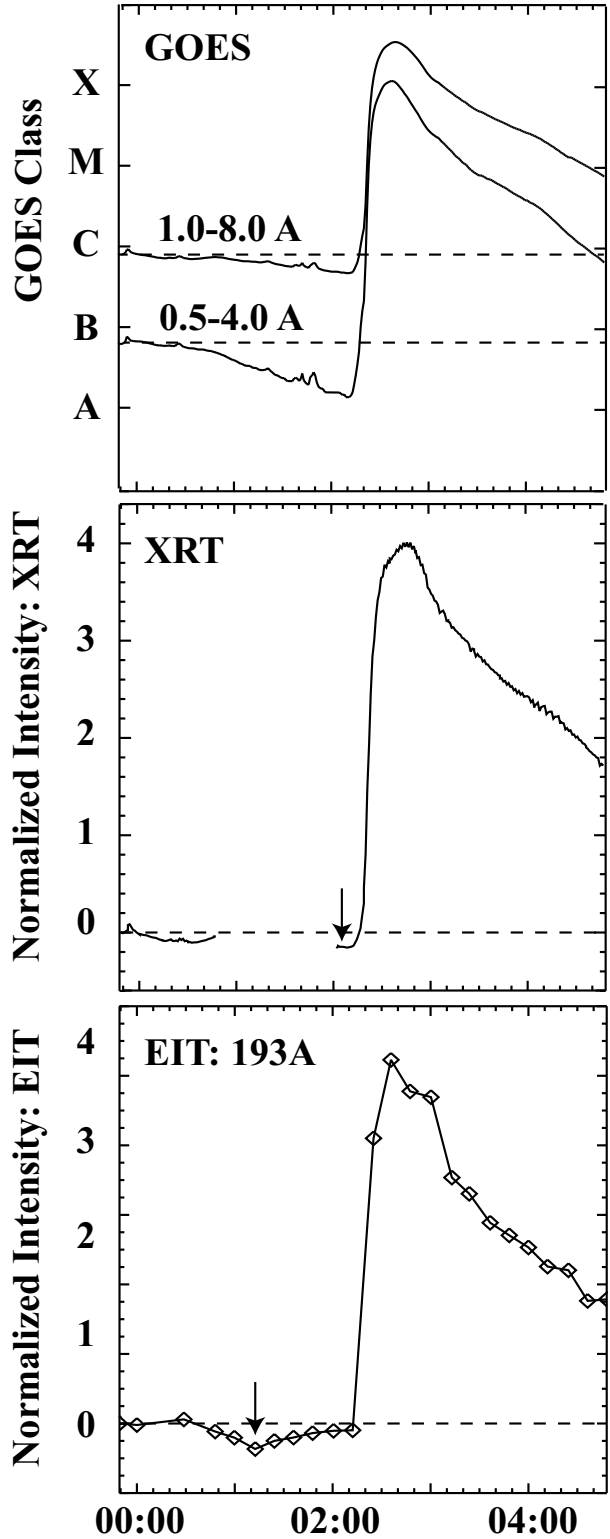}
\end{center}
\caption{ Light curves in soft X-ray and EUV wavelengths. Top panel shows the temporal variation of soft X-ray obtained by GOES. Middle panel shows the XRT soft X-ray intensity integrated over the limited region of Figure 9. Bottom panel shows the EIT 195 \AA~ intensity integrated over the limited region shown in Figure 8. The arrows in the middle and the bottom panel show the time in the right panel of Figures 7 and 6, respectively. 
}\label{test}
\end{figure}

\begin{figure}
\begin{center}
\FigureFile(150mm,80mm){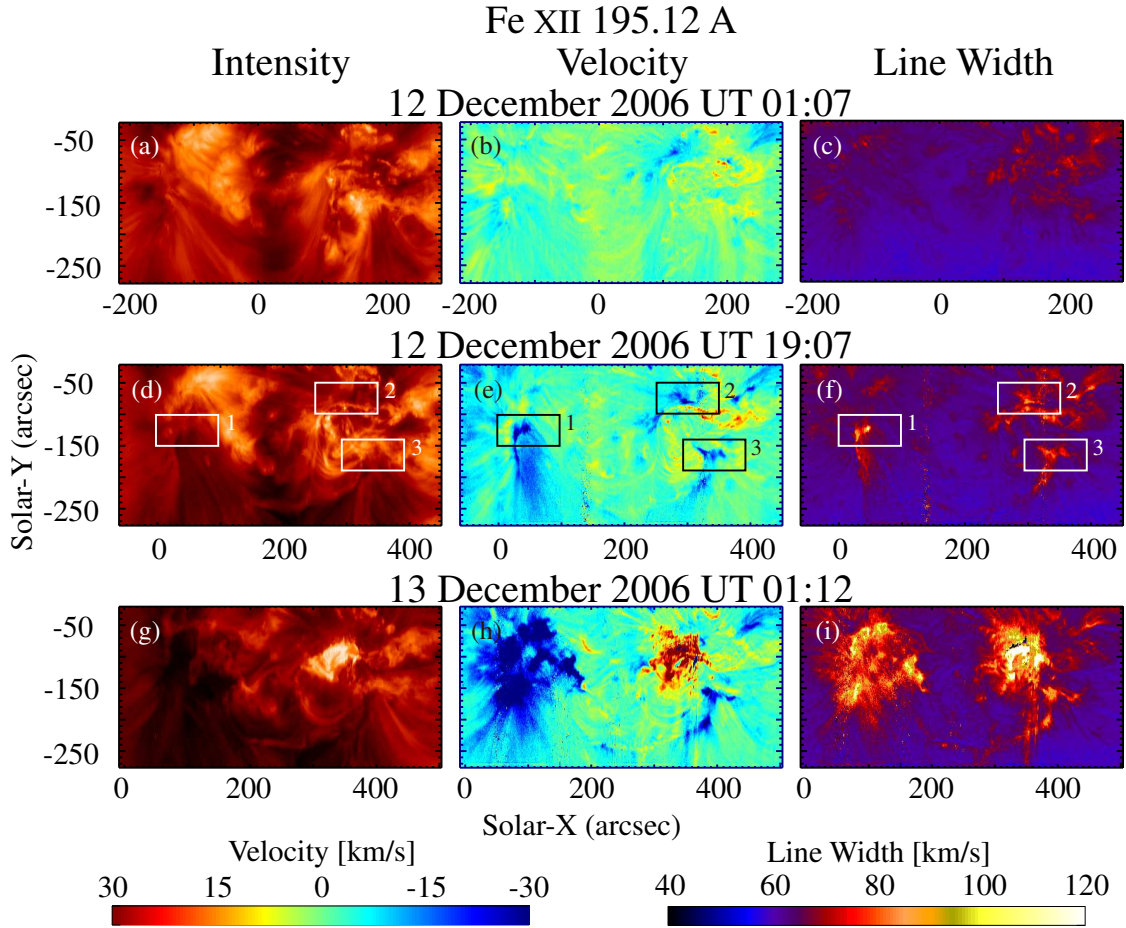}
\end{center}
\caption{ Three sequences of EIS observations (a-c: 01:07 UT on 12th December, d-f: 19:07 UT on 12th December, g-i: 01:12 UT on 13th December). Intensity (a, d, g), velocity (b, e, h), and line width (c, f, i) estimated by Fe\emissiontype{XII} (195.12 \AA).
}\label{test}
\end{figure}

\begin{figure}
\begin{center}
\FigureFile(150mm,80mm){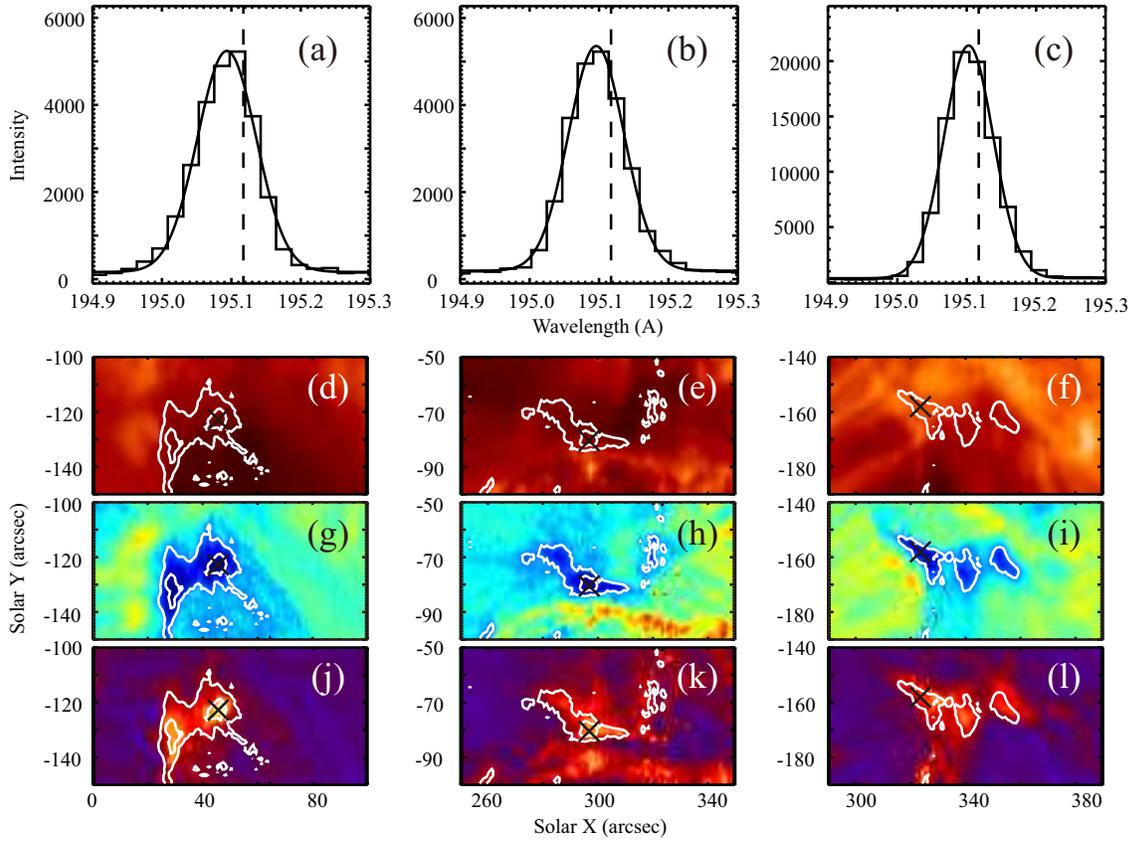}
\end{center}
\caption{ Examples of the line profiles at the blue shifted regions in and around the active region (a-c).
Enlarged display of the EIS observations are shown (d-f: intensity, g-i: velocity, j-l: line width).
White contours represent the blueshift faster than 20 km sec$^{-1}$.
The line spectra shown at the top of the figure corresponds to the x-mark in (d-l).
(a)/(b)/(c) correspond to the x-mark in (d, g, j)/(e, h, k)/(f, i, l), respectively.
}\label{test}
\end{figure}

\begin{figure}
\begin{center}
\FigureFile(150mm,80mm){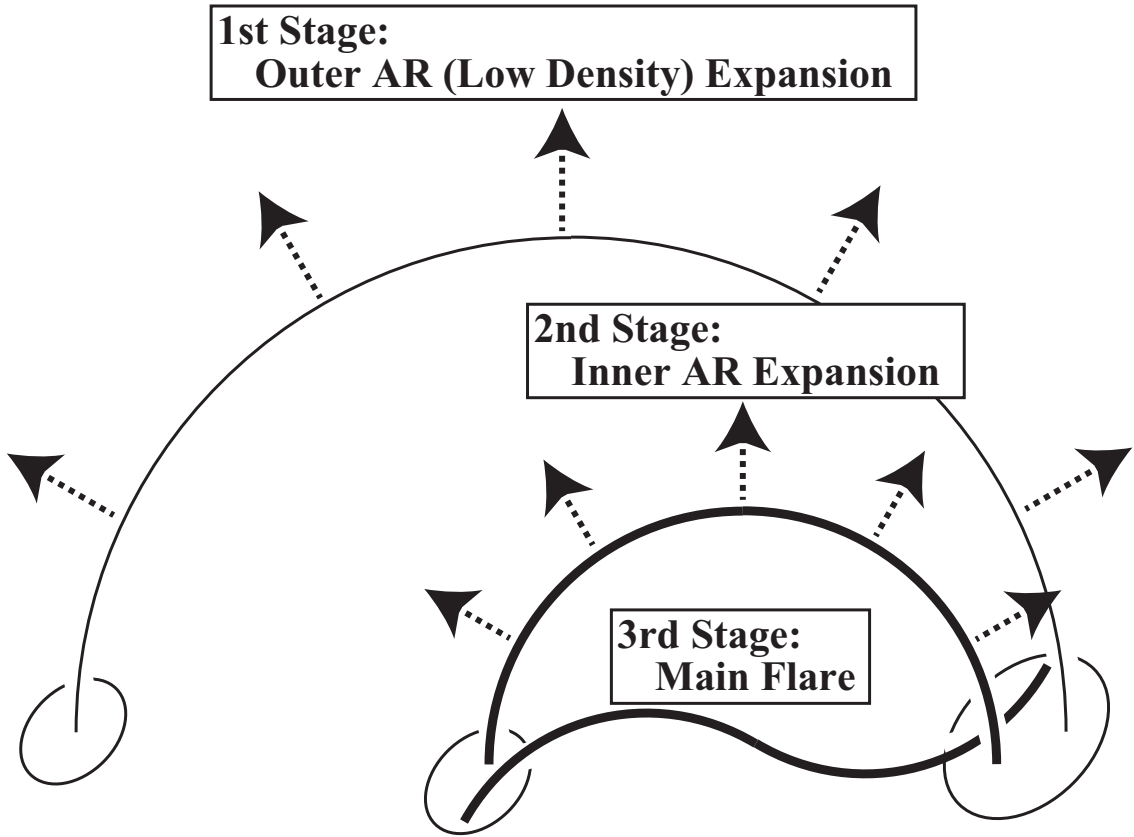}
\end{center}
\caption{ Schematic illustration of the observational results. The arrows represent the expansion of the coronal loops.
}\label{test}
\end{figure}

\end{document}